\documentclass[preprint]{aastex}

\def\cs{{c_{\rm s}}}
\newcommand{\dfrac}[2]{\frac{\displaystyle{#1}}{\displaystyle{#2}}}

\begin{document}

\title{STRUCTURE AND INSTABILITIES OF AN IRRADIATED VISCOUS PROTOPLANETARY DISK}
\author{HIDEKO NOMURA}
\affil{Department of Physics, Graduate School of Science, Kyoto University,
Sakyo-ku, Kyoto 606-8502, Japan; hnomura@tap.scphys.kyoto-u.ac.jp}

\begin{abstract}
We investigate the structure and the stabilities of a protoplanetary
disk, which is heated by viscous process in itself and by
its central star. The disk is set to rotate with the Keplerian velocity
and has the surface density distribution of the minimum mass solar nebula.
We assume the vertical hydrostatic equilibrium and the radiative
equilibrium at each point, and solve the two-dimensional radiative
transfer equation by 
means of the Short Characteristic method in the spherical coordinate 
in order to determine the disk structure. Our calculation shows that
at the outer region of the disk with a distance from the central star
of $x>1$AU the radiative heating from the inner disk dominates the
viscous heating even near the midplane.
It is because of the high temperature distribution in the optically thin
surface layer and the relatively high disk height ($z_{\infty}\sim 0.7x$
at $x\sim 1$AU) as a consequence of the irradiation from the inner
hot region of the disk. In addition, we examine the convective and the
magnetorotational instabilities of the disk. As a result, the whole disk
is convectively stable since the dusty region is not heated by the viscous
dissipation from the midplane but by the radial radiative heating.
On the other hand, almost all the disk is
magnetorotationally unstable except for the region near the equatorial
plane of 2AU $<x<$ 10AU. Finally we discuss the growth and the size
distribution of dust particles in the disk, which suggests that
there exist cm-sized particles in the surface layer,
namely, in the exposed region of the disk.
\end{abstract}


\section{INTRODUCTION}

It has been established that there exist circumstellar disks around a large
fraction of young stellar objects (YSOs), some of which will evolve to
form planetary systems analogous to our Solar System 
(Wilking, Lada, \& Young 1989; Stauffer et al. 1994; Haisch, Lada \&
Lada 2001). Infrared excesses over the photospheric emissions, often
observed in the spectral energy distributions (SEDs) of YSOs, are considered
to be radiated from dusty circumstellar disks (Cohen 1974; Adams \& Shu
1985; Kenyon \& Hartmann 1987; Beckwith \& Sargent 1993; Hartmann 1998).
Furthermore, the disk-structure images with radii of $\sim 100$AU
are sometimes observed around YSOs at optical and near-infrared
wavelengths as silhouettes against the background bright nebula (O'Dell et
al. 1993; O'Dell 1998; Bally, O'Dell \& McCaurean 2000) or as optically
thick dust lanes with optically thin reflection nebulae, absorbing and
scattering light from their central stars (Burrows et al. 1996; Padgett et
al. 1999).  

These circumstellar disks play vital roles on the star and planetary
system formation: for example, they transform angular momentum outward
and accrete mass to their central stars (e.g., Spitzer 1978; Shu, Adams, \&
Lizano 1987; Bodenheimer 1995; Papaloizou \& Lin 1995; Nomura \&
Mineshige 2000; Stone et al. 2000), and they form planets in them
through the various processes of dust grain growth (e.g., Safronov
1969; Weidenschilling \& Cuzzi 1993; Beckwith, Henning \& Nakagawa 2000). 
The turbulence in the disks is one of the keys to elucidate their structures
and evolutions. The turbulent viscosity in differentially rotating disks
transforms the angular momentum and mass, and releases the gravitational
energy of accreted mass as heat through the viscous dissipation (e.g.,
Shakura \& Sunyaev 1973; Lynden-Bell \& Pringle 1974; Pringle 1981). 
A number of mechanisms to excite turbulent motion have been 
investigated for various regions of disks in varied evolutional stages,
such as the thermal convection, the magnetorotational instability, the
shear motion due to differential rotation, the infalling envelope, the
stellar wind, and so forth (e.g., Lin \& Papaloizou 1980; Cabot et
al. 1987; Balbus \& Hawley 1991; Sano et al. 2000; Zel'dovich 1981;
Sekiya 2000; Elmegreen 1978; Cameron 1978), but we do not yet have a
universally accepted scenario.
Meanwhile, many studies have been done on the structure of such optically
thick viscous (active) protostellar disk under the assumptions of the vertical
hydrostatic equilibrium and the energy balance between the viscous
heating and the radiative cooling, sometimes in the context of explaining
the FU Ori outburst phenomena (e.g., Lin \& Papaloizou 1980; Pringle
1981; Lin \& Papaloizou 1985; Kawazoe \& Mineshige 1993; Bell \& Lin 1993). 

In order to reproduce the structure of protoplanetary disk, however,
it is not sufficient to evaluate only the vertical viscous heating from
the midplane since the disks are irradiated from the central star and the
inner hot region of the disk
(e.g., Kenyon \& Hartmann 1987; Chiang \& Goldreich 1997; D'Alessio et
al. 1998; Bell 1999). This irradiation process is noteworthy since it heats up
and flares up the disk surface, reproducing the observed flat SEDs of 
far-infrared radiation from YSOs (the radiation flux of $\nu
F_{\nu}\propto \nu^0$), which is not recreated by radiation from
geometrically thin accretion disks (e.g., Lynden-Bell \& Pringle 1974;
Adams, Lada, \& Shu 1987). The flared features of optically thick disks 
are in fact observed as shadows against scattering starlight in
optically thin circumstellar matter (e.g., Burrows et al. 1996; Padgett et
al. 1999). 
The profiles of such passive disks irradiated only from their central stars
have been investigated under the assumption of the vertical hydrostatic
equilibrium and with a simple model of radiative transfer of starlight,
without and with the turbulent viscous heating (e.g., Kenyon \& Hartmann
1987; Chiang \& Goldreich 1997; Hubeny 1991; D'Alessio et al. 1998). 

In this paper we strictly treat the radiation processes by solving the
two-dimensional radiative transfer equation so as to investigate the
structure and the instabilities of an axisymmetric disk under the
influences of the viscous heating and the reprocessing of radiation from
the central star and the inner hot region of the disk. 
In the next section we present the disk model, which is heated by the
viscous dissipation and its central star, and assumed to satisfy
the vertical hydrostatic equilibrium and the local radiative
equilibrium. 
In \S 3 we calculate the density and
temperature distributions of the disk. The convective and the 
magnetorotational instabilities of the disk, which will excite
turbulent motion, are investigated in \S 4, and the dust size distribution
in the disk is discussed by means of the turbulent-eddy-trapping model
for dust growth in \S 5. Finally we summarize the results in \S 6.

\section{DISK MODEL}

We consider an axisymmetric disk surrounding a central star with 
physical parameters of the typical T Tauri stars; the mass of
$M_*=0.5M_{\odot}$, the radius of $R_*=2R_{\odot}$, and the temperature
of $T_*=4000$K (e.g., Kenyon \& Hartmann 1995).
We investigate the density and temperature distributions of the disk
under the assumptions of the hydrostatic equilibrium in the vertical
direction (\S 2.1) and the radiative equilibrium at each point (\S 2.3).
We also hypothesize that dust and gas have the same temperature
distribution allover the disk in this paper.
For a matter of convenience, we adopt the cylindrical coordinate $(x,z)$
in \S 2.1 and the spherical coordinate $(r,\Theta)$ [where
$r\equiv(x^2+z^2)^{1/2}$ and $\Theta\equiv\tan^{-1}(x/z)$] in \S 2.3 , 
in both of which we put the central star at the origin. 

\subsection{Hydrostatic Equilibrium in the Vertical Direction}

In order to determine the density distribution, $\rho(x,z)$, we assume
the vertical hydrostatic equilibrium,
\begin{equation}
\dfrac{dP}{dz}=\dfrac{\cs^2}{\rho}\dfrac{d\rho}{dz}=-\rho g_z. \label{1}
\end{equation}
The sound speed $\cs$ is defined as $\cs^2\equiv dP/d\rho=kT/m_{\mu}$, where
$P$, $T$, $k$ and $m_{\mu}$ represent the pressure, the temperature, the
Boltzmann's constant, and the mean molecular mass, respectively, and we
adopt $m_{\mu}=2.3m_{\rm H}$ ($m_{\rm H}$ is the hydrogen mass). The
circumstellar disks around T Tauri stars are 
considered to be rotationally supported. Then 
the gravitational force in the vertical direction $g_z$ is written as
$g_z=\Omega^2 z$ with the Kepler frequency,
\begin{equation}
\Omega(x)=(GM_*/x^3)^{1/2}\approx 1.4\times 10^{-7}{\rm s}^{-1}(x/1{\rm AU})^{-3/2}, \label{2}
\end{equation}
where we adopt $M_*=0.5M_{\odot}$ for the mass of the central star.
For the boundary condition, we put $\rho(x,z_{\infty})\equiv 3.8\times
10^{-22}$ g cm$^{-3}$, which corresponds to the typical number density of 
molecular clouds, $n\approx 10^2$ cm$^{-3}$ (where the mean molecular mass
of $m_{\mu}=2.3m_{\rm H}$ is assumed). The height of the disk $z_{\infty}$
is determined by the condition,
\begin{equation}
\int_{-z_{\infty}}^{z_{\infty}}\rho(x,z)dz=\Sigma(x), \label{3}
\end{equation}
where the surface density of the disk $\Sigma$ is put to agree with that of the minimum
mass solar nebula (e.g., Hayashi, Nakazawa, \& Nakagawa 1985), 
\begin{equation}
\Sigma(x)=1.7\times 10^3\ {\rm g\ cm}^{-2}(x/1{\rm AU})^{-3/2}. \label{4}
\end{equation}
Solving equation (\ref{1}) with the above conditions and the temperature
distribution $T(x,z)$, which is derived in subsection \S 2.3, by means of the
fourth-order Runge-Kutta method, we get the density distribution $\rho(x,z)$.

\subsection{Heating Sources}

For the heating sources of the disk we consider two kinds of processes;
the viscous dissipation in itself and the gravitational energy release
associated with the contraction of its central T Tauri star.
In protoplanetary disks it is believed that some kinds of instabilities
(see \S 4) cause the turbulent motion and then induce the angular
momentum and mass transfer. This process is accompanied with the heating
of the disks via viscous dissipation of turbulence. Following the
so-called $\alpha$-viscous model, we represent it by putting heating
source at the midplane of the disk (at $z=0,\Theta=\pi/2$) with a
heating rate of 
\begin{equation}
Q^+_{\rm vis}=(9/4)\Sigma\alpha\cs_0^2\Omega, \label{5}
\end{equation}
where $\cs_0$ is the sound speed at the midplane.
In this model the kinetic viscosity is prescribed as
$\nu_{\rm vis}=\alpha\cs_0 H=\alpha\cs_0^2/\Omega$, where $H$ is the
scale height of the disk
(e.g., Shakura \& Sunyaev 1973; Pringle 1981; Kato, Fukue \& Mineshige
1998), and $\alpha=10^{-2}$ is adopted here. The vicinity of the
midplane at the inner region of the disk is mainly heated by
diffusive radiation from this viscous heating source as we will see in
\S 3. Besides,
we put thermal radiation source at the stellar surface
(at $r=R_*$) with a heating rate of  
\begin{equation}
Q^+_{\rm star}=\sigma T_*^4, \label{6}
\end{equation}
where $\sigma$ is the Stefine-Boltzmann constant, and examine the
reprocessing of radiation from the central star in the disk by solving
the two-dimensional radiative transfer equation as we describe in the
following subsection (\S 2.3) and Appendix. For the stellar temperature
$T_*=4000$K is adopted. We here comment that the irradiation not only
from the central star but also from the inner hot disk on the surface
region of the outer disk is automatically simulated in our calculation,
and the latter dominates the former in our model as we will see in \S 3. 

\subsection{Radiative Equilibrium}

Evaluating the temperature distribution $T(r,\Theta)$, we assume
the radiative equilibrium (that is, the emitted and absorbed
radiation are balanced) at each point $(r,\Theta)$ 
[$r\equiv(x^2+z^2)^{1/2}$ is the distance from a central star and
$\Theta\equiv\tan^{-1}(x/z)$ is the angle from the $z$-axis] as
\begin{equation}
4\pi{\displaystyle\int_0^{\infty}\kappa_{\nu}(r,\Theta) B_{\nu}[T(r,\Theta)]d\nu=\int_0^{\infty}\kappa_{\nu}(r,\Theta)J_{\nu}(r,\Theta)d\nu}, \label{7}
\end{equation}
where $\nu$ is the photon frequency, $B_{\nu}(T)$ the Plank function for
blackbody radiation, and $\kappa_{\nu}$ the monochromatic
opacity given by the model in the following subsection (\S 2.4).
In equation (\ref{7}) local thermodynamics equilibrium
$\eta_{\nu}(r,\Theta)=\kappa_{\nu}(r,\Theta) B_{\nu}[T(r,\Theta)]$ is assumed,
where $\eta_{\nu}$ is the monochromatic emissivity (e.g., Mihalas 1978).
The mean intensity $J_{\nu}(r,\Theta)$ is given by integrating 
the specific intensity $I_{\nu}(r,\Theta;\mu,\phi)$ of radiation field
for all solid angles $d\omega\equiv d\mu d\phi$ as
\begin{equation}
J_{\nu}(r,\Theta)=\dfrac{1}{4\pi}\int_0^{2\pi}\int_{-1}^1 I_{\nu}(r,\Theta;\mu,\phi)d\mu d\phi, \label{8}
\end{equation}
where $\mu$ is the cosine of the angle between the ray direction
and the vector $\overrightarrow{OP}$ ($O=[0,0], P=[r,\Theta]$), and
$\phi$ is the angle between the projection of the ray to the plane whose
normal is $OP$ and the line ($\phi=0$) on the plane parallel to the
midplane of the disk ($z=0,\Theta=\pi/2$) (cf. Fig. 3 of Dullemond \&
Troulla 2000). We here divide the total specific intensity $I_{\nu}$
into three parts as
\begin{equation}
I_{\nu}=I^*_{\nu}+I^{\rm vis}_{\nu}+I^{\rm th}_{\nu},
\end{equation}
where $I^*_{\nu}$ and $I^{\rm vis}_{\nu}$ represent radiation directly
from the heating sources described in \S 2.2, the central star and the
viscous dissipation at the midplane of the disk, respectively, and
$I^{\rm th}_{\nu}$ arises from thermal emission of dust grains in the
disk (e.g., Shu 1991; Efstathiou \& Rowan-Robinson 1990). We note that
we neglect the radiative scattering for the simplicity in this paper.
Each of them is derived from radiative transfer equation,
\begin{equation}
\dfrac{dI^s_{\nu}}{ds}=\rho\kappa_{\nu}(S^s_{\nu}-I^s_{\nu}), \label{9}
\end{equation}
where the source functions $S^*_{\nu}=S^{\rm vis}_{\nu}=0$ and $S^{\rm
th}=B_{\nu}$. From the above equation, the intensity $I^*_{\nu}$ and
$I^{\rm vis}_{\nu}$ are derived with $T_*=T(r=R_*,\Theta)$ and $T_{\rm
vis}(x)=T(r,\Theta=\pi/2)$ as
\begin{equation}
I^*_{\nu}(r,\Theta;\mu,\phi)=B_{\nu}(T_*)e^{-\tau_{\nu}(r,\Theta)} \label{21}
\end{equation}
for $\mu<-(r^2-R_*^2)^{1/2}/r$ [that is, $(\mu,\phi)$ originates from
the central star] and $I^*_{\nu}(r,\Theta;\mu,\phi)=0$ otherwise, and
\begin{equation}
I^{\rm vis}_{\nu}(r,\Theta;\mu,\phi)=B_{\nu}[T_{\rm vis}(r)]e^{-\tau_{\nu}(r,\Theta)}
\end{equation}
when $(\mu,\phi)$ originates form the midplane
$(r,\Theta=\pi/2)=(x,z=0)$ and $I^{\rm vis}_{\nu}(r,\Theta;\mu,\phi)=0$
otherwise. The viscous heating temperature $T_{\rm vis}(r)$ is related
with $\cs_0$ in equation (\ref{5}) as $\cs_0^2=kT_{\rm vis}/m_{\mu}$. 
The intensity $I^{\rm th}_{\nu}$ is derived by integrating equation
(\ref{9}) along a ray, $s$ (which is directed to $[\mu,\phi]$), as
\begin{equation}
I^{\rm th}_{\nu}(r,\Theta;\mu,\phi)=\int_0^s \kappa_{\nu}(r',\Theta')\rho(r',\Theta')B_{\nu}[T(r',\Theta')]e^{-\tau_{\nu}(r',\Theta')}ds'. \label{10}
\end{equation}
In equations (\ref{21})-(\ref{10}) $\tau_{\nu}(r,\Theta)$ represents the
specific optical depth at the point $(r,\Theta)$ from the radiation
source $(r',\Theta')$, given by
\begin{equation}
\tau_{\nu}(r,\Theta)=\int_0^{s}\kappa_{\nu}(r',\Theta')\rho(r',\Theta')ds'. \label{11}
\end{equation}
Together with 
the density distribution $\rho(r,\Theta)$ obtained in \S 2.1, we solve
equations (\ref{7})-(\ref{11}) self-consistently to get the temperature
distribution $T(r,\Theta)$ in the disk. The numerical method to solve
equation (\ref{7}) is described in Appendix A.1, and that to calculate
equations (\ref{21})-(\ref{11}) is in A.2. We note that we do not
consider energy transport by convection here, but it turns out in \S 4.1
that this is permissible since the whole of the disk is in
fact convectively stable in our disk model. 

\subsection{Opacity}

In this paper we consider that the disk consists of dusty medium (in
which dust is well mixed with gas) at low temperature ($T<2300$K) and
only of gaseous matter at $T>2300$K. For dusty medium we use the frequency
dependent absorption coefficient $\kappa_{\nu}$ of Adams \& Shu (1985, 1986;
see also Draine \& Lee 1984). In their model the dust is constituted by
three types of components of graphite and 
olivine silicate grains, and water icy mantle which coats the grains.
They are assumed to evaporate at the characteristic dust destruction 
temperatures of 2300 K (graphite), 1500 K (silicate), and 150 K (ice).
For gaseous matter we provisionally take the weighted transmission
averaged opacity, the Rosseland mean opacity $\kappa_{\rm R}$, defined as 
$1/\kappa_{\rm R}\equiv\int_0^{\infty}(1/\kappa_{\nu})(\partial B_{\nu}/\partial T)d\nu/\int_0^{\infty}(\partial B_{\nu}/\partial T)d\nu.$
The density and temperature dependent model of Bell \& Lin (1994),
$\kappa_{\rm R}=\kappa_i\rho^aT^b$, is adopted here, where $\kappa_i$,
$a$ and $b$ are given for each of eight states of gas. It represents the
Alexander/Cox/Stewart opacity and covers the wide temperature range
of $10^3{\rm K}<T<10^7$K (see Bell \& Lin 1994 in details and see also
Lin \& Papaloizou 1985). The applicability of this approximate
treatment of the opacity for gas is discussed in \S 3.


\section{DENSITY AND TEMPERATURE PROFILES OF THE DISK}

Solving the equations of the vertical hydrostatic equilibrium (\ref{1})
and the local radiative equilibrium (\ref{7}) iteratively together with
the two-dimensional radiative transfer equation (\ref{9}) as we
mentioned in the previous section and Appendix, we obtain the
self-consistent density and temperature distributions of the disk
heated by the viscous process in itself and by its central star. Figure 1a
shows the contour plots of the resultant
temperature ({\it solid lines}) and density ({\it dotted lines}) profiles 
in the $x$-$z$ plane of the `irradiated disk' [which means that it is
the result of the two-dimensional calculation of radiative transfer].
The contour levels are $T=10,10^2,10^3,10^4$, and $10^5$K for the
temperature, and $\rho=10^{-18},10^{-16},10^{-14},10^{-12},10^{-10}$,
and $10^{-8}$g cm$^{-3}$ for the density. We also plot the vertical
temperature and density distributions at
$x=0.046,0.1,0.21,0.46,1.0,2.1,4.6,10$, and $21$AU in 
Figure 2a and 3a, respectively. We here mention the applicability
of our opacity model for gas ($T>2300$K). Whereas we use the Rosseland
mean opacity in this paper as we stated in \S 2.4, the Planck
mean opacity is to be used in equation (\ref{10}) and the left side of
equation (\ref{7}), and the energy-weighted one in the right side of
equation (\ref{7}) (see Hubeny 1990). This discrepant treatment of the
opacity will modify the temperature profile at the optically thin region
of the inner disk because the difference among these opacities goes up to
several orders of magnitude at the maximum (e.g., Alexander \& Ferguson
1994). The temperature near the midplane is, however, hardly affected
since it is determined by equation (\ref{12}) as we stated in Appendix.
For comparison we display those of the `non-irradiated disk' [which is
derived from the energy transport equation only in the vertical
direction (\ref{12})] 
in Figure 1b, 2b, and 3b in the same manner as Figure 1a-3a.
The constant temperature near the surface in Figure 1b and 2b is
due to our inappropriate use of the diffusion approximation (eq. [A1])
even at the optically thin region.

The remarkable differences between the `irradiated' and `non-irradiated' 
cases appear in the rises of temperature at the surface region and at
the equatorial plane outside $x\sim 1$AU of the `irradiated disk'. As a
result, the vertical height of the `irradiated disk' becomes higher than
that of the `non-irradiated disk'. The effect of this flaring is due to
the irradiation from the inner region and conspicuous at the outer disk,
as is modeled in Kenyon \& Hartmann (1987), especially at the optically
thin surface layer (Chiang \& Goldreich 1997; D'Alessio et al. 1998). 
The main source of the irradiation is, however, the inner hot disk rather
than the central star in this case, contrasting with the previous works. 
We next discuss the reason for the radiative heating from the inner disk
to dominate the viscous heating at $x> 1$AU. 
Now the energy density due to the viscous heating at the midplane of the disk
decreases with radius approximately as 
$E_{\rm vis}=aT_{\rm vis}^4\propto r^{-10}$,
which is derived from the equations of the energy balance and the
radiative transfer in the vertical direction,
$(9/4)\Sigma\alpha\cs_0^2\Omega=\sigma T_{\rm
eff}^4\sim(4/3)acT_{\rm vis}^4/\kappa_{\rm R}\Sigma$, 
where we use $\Sigma\propto r^{-3/2}$, $\Omega\propto r^{-3/2}$,
$\cs_0^2\propto T_{\rm vis}$, and $\kappa_{\rm R}\propto T_{\rm vis}^{6/5}$ 
(dust grains mainly contribute to the opacity in this region). 
Next let us consider the radial dependence of the energy density due to
the radiative heating from the inner disk.
If we postulate the extreme case
that the height was large enough to adopt the plane-parallel
approximation in the radial direction, the dependence would became 
$E_{\rm rad}=aT_{\rm rad}^4\sim(3F_{\rm rad}/4c)
\int\kappa_{\rm R}\rho dr\propto r^{-5}$, where $F_{\rm rad}(={\rm
const.})$ is the radial radiative flux and we use 
$\kappa_{\rm R}\propto T_{\rm rad}^{6/5}$ and $\rho\propto r^{-5/2}$
from our calculation. In this case the radiative heating would surely
dominate the viscous heating. 
On the contrary, if the disk is geometrically thin enough, the radiative
heating from the inner disk becomes negligible. Thus one of the reasons
for our result is the high vertical height of the disk.
In addition the temperature at the surface layer is higher than that at
the equatorial plane, which allows the effect of radiative heating from
the inner region stronger.

In conclusion our results show that the surface layer of the disk is
heated up and flared via the irradiation from the inner hot disk, and
that the radiative heating from the inner disk dominates the viscous 
heating at $x> 1$AU because of the high disk height and the high
temperature at the disk surface in this case.


\section{DISK INSTABILITIES}

In protoplanetary disks it is believed that some kinds of instabilities
cause the turbulent motion, and then, induce the angular momentum and mass
transfer, and/or the energy transport. A number of processes have been
proposed, among which we examine in this section two
of intrinsical mechanisms in the disks before the dust settling stage:
the convective instability (\S 4.1) and the magnetorotational
instability (\S 4.2), making use of the temperature and density
profiles obtained in the previous section.

\subsection{Convective Instability}

It has been suggested that the protoplanetary disks are convectively
unstable because they consist of dusty components whose opacity has a 
temperature dependence of $\kappa_{\rm R}\propto T^{\beta}$, $\beta\geq 1$,
and then causes a superadiabatic gradient against the direction of
gravitational force (Lin \& Papaloizou 1980, see also Cameron \& Pine 1973;
Ruden, Papaloizou, \& Lin 1988; Kley, Papaloizou, \& Lin 1993;
Stone \& Balbus 1996). Now we examine the convective instability of the
disk, which has the temperature and density distributions shown in \S 3,
using the criterion of
\begin{equation}
\Delta_z-\Delta_{\rm ad}>0,\ \ \ \Delta_z\equiv\dfrac{\partial\log T}{\partial\log P},\ \ \ \Delta_{\rm ad}\equiv\dfrac{\gamma-1}{\gamma}, \label{13}
\end{equation}
where $\gamma$ is the specific heat ratio and the pressure satisfies
$P=\gamma\cs^2\rho$ (e.g., Cox 1980). We here take $\gamma=7/5$ since 
the main component of the disk is molecular hydrogen. 
As a result, we find that the whole of the `irradiated disk' [obtained
from the two-dimensional calculation of radiative transfer] is
convectively stable, 
while the dusty region ($T<2300$K) of the `non-irradiated disk' [derived
from only the vertical radiative transfer] is unstable 
(Fig. 4b, {\it dotted stripe} region).  
The unstable region disappears in the `irradiated disk' because its
dusty region is not heated from the midplane but from the inner
region as we see in \S 3.

\subsection{Magnetorotational Instability}

Magnetorotational instability is one of the promising turbulent sources
in accretion disks (Velhikov 1959; Balbus \& Hawley 1991; 
Balbus \& Hawley 1998 and references therein). When a differentially
rotating, fully ionized, magnetized disk satisfies the condition, 
\begin{equation}
2\pi v_{\rm A}/\Omega\leq H, \label{14}
\end{equation}
the disk is magnetorotationally unstable,
where $\Omega$ and $H$ are the rotational frequency and the disk scale
height, respectively, and $v_{\rm A}$ is the Alfv\'{e}n speed, defined
with the magnetic field $B$ and the density $\rho$ as $v_{\rm A}\equiv
B/(4\pi\rho)^{1/2}$ (Balbus \& Hawley 1991). In protoplanetary disks, on
the other hand, the ionization degree of material is low enough (e.g.,
Umebayashi \& Nakano 1988) that the effects of the ohmic dissipation and
the ambipolar diffusion should be taken into account (e.g., Blaes \&
Balbus 1994; Jin 1996; Gammie 1996; Sano et al. 2000).
Here we examine the magnetorotationally unstable region in the disk with
the temperature and density distributions obtained in \S 3 under the
assumption that the ratio $\cs_0/v_{\rm A}=10$ allover the disk.
According to Sano et al. (2000), in which the stabilizing effect of the
ohmic dissipation in protoplanetary disks is investigated, the disk is
magnetorotationally unstable when 
\begin{equation}
2\pi\max[v_{\rm A}/\Omega,\eta/v_{\rm A}]\leq H, \label{15}
\end{equation}
where $\eta$ is the magnetic diffusivity. We here consider only the
collisions between electrons and neutrals, mainly hydrogen molecules and
helium atoms, as contributor to the diffusivity $\eta$ [the
contributions of ions and charged grains are negligible at $x\geq 1$AU
(Sano et al. 2000)]. Thus $\eta$ is given by
\begin{equation}
\eta\equiv c^2/4\pi\sigma_c=(c^2m_e{\rm <}\sigma v{\rm >}_e/4\pi e^2)(n/n_e), \label{16}
\end{equation}
where $c$, $\sigma_c$, $m_e$, $e$, $n$, and $n_e$ are the light speed,
the electrical conductivity, the electron mass and charge, the number
densities of neutrals and electrons, respectively. 
For the momentum transfer rate between electrons and neutrals, 
${\rm <}\sigma v{\rm >}_e$, we use the experimental formula in Appendix
of Sano et al. (2000; see also Hayashi 1981). 
Meanwhile, we evaluate the ionization degree $\xi\equiv n_e/n$ at
$T\leq 10^3$K with the ionization rate $\zeta$, the temperature $T$, and
the molecular number density $n$ as
$\xi=3.4\times 10^2T^{1/4}\zeta^{1/2}n^{-1/2}$
(Gammie 1996), where $\zeta$ is represented as
$\zeta=\zeta_{\rm CR}\exp(-\tau_{\rm CR})+\zeta_{\rm R}$.
Each of $\zeta_{\rm CR}$ and $\zeta_{\rm R}$ is the
ionization rate by cosmic rays and radioactive elements, 
respectively. Following Umebayashi \& Nakano (1980), we adopt
$\zeta_{\rm CR}=10^{-17}$s$^{-1}$ and $\zeta_{\rm R}=6.9\times
10^{-23}$s$^{-1}$.
The ionization depth by cosmic rays, $\tau_{\rm CR}$, is written as
$\tau_{\rm CR}=\int_z^{\infty}\rho(x,z)dz/\chi_{\rm CR}$,
where $\chi_{\rm CR}$ is the attenuation length and we use $\chi_{\rm
CR}=96$g cm$^{-2}$.
At $T>10^3$K, where the thermal collision dominates the cosmic rays and
the radioactive elements on ionization, we adopt the thermal ionization
degree $\xi$ of Umebayashi (1983). In addition,
in order to see the effect of ambipolar diffusion, we examine the ratio
of the collision frequency between ions and neutrals to the epicyclic
frequency, $\gamma_{\rm d}\rho_{\rm i}/\Omega$,
where $\gamma_{\rm d}\approx 3.5 \times 10^{13}$ cm$^3$ g$^{-1}$
s$^{-1}$ is the drag coefficient (e.g., Shu 1983) and $\rho_{\rm i}$ is
the ion density (Blaes \& Balbus 1994; Hawley \& Stone 1998). 

In Figure 4a we plot the resultant magnetorotationally unstable region
({\it solid stripe}) in the $x$-$z$ plane of the `irradiated disk', whose
density and temperature distributions are obtained from our
two-dimensional 
numerical calculation. The figure shows that almost all the disk is 
magnetorotationally unstable except for the region near the equatorial
plane of 2AU $<r<$ 10AU. The ionization degree is high enough for the 
disk to be unstable in the inner region because of the thermal
collisions, and in the surface and outer regions owing to the cosmic rays.
Equation (\ref{15}) suggests that the stable region spreads as
$\cs_0/v_{\rm A}$ becomes larger and vice versa (although we here
examine only the case of $\cs_0/v_{\rm A}=10$) since the relation
$v_{\rm A}/\Omega<\eta/v_{\rm A}$ is satisfied allover the disk in this
case (see also Sano et al. 2000).
The stabilization due to the ambipolar diffusion is ignorable
in this case ($\gamma_{\rm d}\rho_{\rm i}/\Omega\leq 10^{-2}$ in allover
the region which the ohmic dissipation dose not stabilize; see
Stone \& Hawley 1998). For comparison we also plot the unstable region
of the `non-irradiated disk', which is obtained from the
energy transport equation only in the vertical direction, in Figure 4b. 
The stable region in the `irradiated disk' is smaller than that in the
`non-irradiated disk' because of its hotter temperature profile. 
We note that our assumption that the disk is heated by viscous
dissipation at every radius (\S 2.2) is conflict with the fact that the
stable region exists in 2AU $<r<$ 10AU. But our results will not be
affected by this inconsistency since this region is heated not viscously
but radiatively as we show in \S 3.

In summary we conclude that the turbulent source in a protoplanetary disk 
is not convective instability but magnetorotational instability under the
conditions that the disk has the surface density distribution of minimum
mass solar nebula and the viscous parameter of $\alpha=10^{-2}$.

\section{DUST SIZE DISTRIBUTION}

Protoplanetary disks are considered to consist at first of a few
$\times 10^{-1}\mu$m sized dust grains, originating from the
interstellar medium, which grow larger particles and at last lead to the
planet formation through the processes of sticking, gravitational
attraction, and gas accretion (e.g., Safronov 1969; Weidenschilling \&
Cuzzi 1993; Beckwith, Henning \& Nakagawa 2000 and references therein).
For the dust growth process numerous theoretical models have been proposed, 
for example, the collisional models in the laminar nebula and in the
turbulent accretion disk, and the particle concentration model owing to
turbulent motion (e.g., Kusaka, Nakano, \& Hayashi 1970; Weidenschilling
1980; Nakagawa, Nakazawa, \& Hayashi 1981; Mizuno 1989; Squires \& Eaton
1991; Klahr \& Henning 1997; Supulver \& Lin 2000). On the other hand,
the evidences of dust grain growth in protoplanetary disks are
observationally presented by means of the changes in their spectral
energy distributions (e.g., Beckwith \& Sargent 1991; Beckwith et
al. 2000; D'Alessio, Calvet, \& Hartmann 2001; Throop et al. 2001).
In this section we discuss the dust size distribution in the disk,
which has the density and temperature structures in \S 3 and the turbulent
region obtained in \S 4 for the `irradiated disk', 
dealing with the turbulent-eddy-trapping model of Klahr \& Henning (1997).

Dust particles can be trapped in a turbulent eddy in an accretion disk
when the dust-gas friction time $\tau_{\rm f}$ satisfies the condition,
\begin{equation}
\tau_{\rm f}<l\omega\min(1/g_z,1/2\Omega\Delta V), \label{17}
\end{equation}
where $l$ and $\omega$ are the size and the rotational frequency of the
turbulent eddy, respectively, and $\Omega$ is the orbital frequency
around the central star. $g_z=\Omega^2z$ is the vertical
gravitational force and $\Omega\Delta V$ is the drag force caused by the
difference in orbital velocity, $\Delta V$, between gaseous matter,
undergoing the radial gas pressure, and the Keplerian rotating dust
particles (e.g., Weidenschilling 1977). 
The dust-gas friction time $\tau_{\rm f}$ is given by 
\begin{equation}
\tau_{\rm f}=a_{\rm p}\rho_{\rm d}/\cs\rho, \label{18}
\end{equation}
where $a_{\rm p}$, $\rho_{\rm d}$, $\cs$, and $\rho$ are
the radius and the density of dust particle, the thermal velocity and the
density of gas, respectively. From equations (\ref{17}) and (\ref{18}) we
can estimate the maximum size of dust grains, that can be trapped in
turbulent eddies in the disk, as
\begin{equation}
a_{\rm p,max}=\cs\rho z_{\infty}\Omega\min(1/g_z,1/2\Omega\Delta V), \label{19}
\end{equation}
where we take $l=z_{\infty}$ (disk height), $\omega=\Omega$, and
$\rho_{\rm d}=1$g cm$^{-3}$. In Figure 5 we plot the contour line of the
maximum dust size distribution in the $x$-$z$ plane of the `irradiated
disk', which is estimated from equation (\ref{19}) for the turbulent
(magnetorotationally unstable) and dusty (with temperature of $T<2300$K)
region of the disk. The contour levels are $a_{\rm
p,max}=10^{-2},10^{-1},1,10$, and $10^2$cm and the thick line displays
$z=z_{\infty}$. The figure shows that dust particles can grow to be
cm-size in the surface layer, which is because both of the temperature
and the disk height are high enough there. These particles are
expected to be under the influences of the outside activities, such as x-ray
radiations, the FU Ori outbursts, and so on, which may lead to the
chondrule and/or CAI formations (e.g., Jones et al. 2000).
The dust size growth will also be connected
with the opacity, the structure and the instabilities of the disk. We will
consider these effects and the influence to the spectral energy
distribution of the radiation from the disk in the next paper.

\section{SUMMARY}

In this paper we have evaluated the density and temperature structures
of a protoplanetary disk with the surface density of the minimum mass
solar nebula, rotating with the Keplerian velocity around a central
star, which has the typical parameters of T Tauri stars, and under the
influences of the heating processes via viscosity and the star.
The so-called $\alpha$-model with $\alpha=10^{-2}$ is used here for
turbulent viscosity. 
Assuming the vertical hydrostatic equilibrium and the local radiative
equilibrium, and utilizing the numerical calculation of the
two-dimensional radiative transfer equation, we have obtained the
following conclusions: 

1. The surface layer of the disk is heated and flared up via the
irradiation from the inner hot disk rather than the central star in our
disk model. 

2. At the outer region, $x>1$AU, the radiative heating from the inner
disk dominates the viscous heating because the temperature at the
surface region is high and the disk is geometrically thick, owing to 
the irradiation from the inner hot region of the disk.  

Making use of the resultant density and temperature profiles, we have
examined the convective and the magnetorotational instabilities of the 
disk, which are expected to induce the turbulent motion, and then, bring
about the angular momentum and mass transfer, and/or the energy
transport. As a result, we have found that:

3. The whole disk is convectively stable since 
the dusty region is not heated by the viscous dissipation from the
midplane but the radiative heating from the inner disk. 

4. Almost all of the disk is magnetorotationally unstable as the
ionization degree is high enough that the stabilizations by ambipolar
diffusion and ohmic dissipation are negligible, except for the region
near the midplane of 2AU $<x<$ 10AU. 

Turbulent eddies, excited by magnetorotational instability, will trap dust
grains to facilitate their growth under some conditions. 
We have discussed the size distribution of dust particles in the disk,
which suggests that:

5. The dust particles can grow to be cm-size in the surface layer region
because of the high temperature distribution and the large disk
height. These particles are expected to be exposed to the outside
activities, such as x-ray radiations, the FU Ori outbursts, and so on, 
which may affect the dust grain evolutions.

The influences of the evolution of turbulent motion, 
the dust size distribution, and the surface density
distribution on the structure, the instabilities, and the spectral
energy distribution of the disk should be studied in the next work. 

\acknowledgments

I am grateful to the referee, Dr. I. Hubeny, for his valuable
comments, which improved the clarity of our discussions.
I would like to thank Dr. T. Nakamoto, Dr. T. Sano, and Dr. E.I. Chiang
for very useful comments. I am also appreciative to Dr. S. Inutsuka,
Dr. S. Mineshige, and Dr. H. Kamaya for continuous encouragement. 
This work is financially supported by Research Fellowships of the Japan
Society for the Promotion of Science for Young Scientists, 03055. 

\appendix

\section{Numerical Method}
\subsection{Iterative Method to Calculate Temperature and Density Profiles}

At first we note how we evaluate iteratively the temperature and density
distributions. In the following $T^i(r,\Theta)$ and $\rho^i(x,z)$
represent the results of the $i$-th iterative calculation
of the processes (II)-(V), while $T^j(r,\Theta)$ is
the result of the $j$-th iteration of (a)-(f). The
iterative methods are as follows: 

(I) Initially put the density and temperature profiles of optically
thick accretion disk, $\rho^0(^{\forall}x,^{\forall}z)$ and
$T^0(^{\forall}r,^{\forall}\Theta)$. They are
derived in the conventional manner from the hydrostatic equation
(\ref{1}) in \S 2.1 and the energy transfer equation in the
vertical direction via diffusion (e.g., Lin \& Papaloizou 1980, 1985),
\begin{equation}
\dfrac{4c}{3\kappa_{\rm R}\rho}\dfrac{d(aT^4)}{dz}=\sigma T_{\rm eff}^4, \label{12}
\end{equation}
where $a$ is the radiative constant, defined with the
Stefine-Boltzmann constant $\sigma$ and the light speed $c$ as
$a=4\sigma/c$. In order to solve equation (\ref{12}) we use a
condition $(9/4)\Sigma\alpha\cs_0^2\Omega=\sigma T_{\rm eff}^4$, where
$\cs_0^2=kT(x,z=0)/m_{\mu}$ and $T_{\rm eff}=T(x,z=z_{\rm p})$. 
$z_{\rm p}$ is the photospheric disk height where
the optical depth from the disk surface $z_{\infty}$ becomes 2/3, that
is, $\tau_z=\int_{z_{\infty}}^{z_{\rm p}}\kappa_{\rm R}\rho dz=2/3$.
Here $\kappa_{\rm R}$ is the Rosseland mean opacity described in \S 2.4.
The condition means the energy balance in the vertical direction
between the viscous heating at the midplane and the blackbody radiative
cooling at $z=z_{\rm p}$ (e.g., Meyer \& Meyer-Hofmeister 1982).
These initial density and temperature 
distributions are plotted and used in \S 3 and \S 4 as `non-irradiated
disk' (which means that only vertical energy transport is took into
account), and compared with the `irradiated disk' (obtained from our
two-dimensional radiative transfer calculation below).

(II) Evaluate the temperature distribution 
$T^i(^{\forall}r,^{\forall}\Theta)$ by using
$\rho^{i-1}(^{\forall}x,^{\forall}z)$ and iteratively solving equations
(\ref{7})-(\ref{11}) as we mentioned in \S 2.3: (a) Calculate the
intensity at each point $I_{^{\forall}\nu}(^{\forall}r,^{\forall}\Theta;^{\forall}\mu,^{\forall}\phi)$
by using tentative temperature distribution
$T^j(^{\forall}r,^{\forall}\Theta)$ and integrating equation (\ref{10})
with the Short Characteristics method described in \S A.2.
(b) Substitute the intensity at a point 
$I_{\nu}(r,\Theta;^{\forall}\mu,^{\forall}\phi)$ into equation (\ref{8})
to obtain the mean intensity $J_{\nu}(r,\Theta)$ at that point. (c)
Evaluate the corrected temperature $T^{*j}(r,\Theta)$ at that point from
equation of radiative equilibrium (\ref{7}). (d) Estimate the
temperature correction 
in context of the Newton-Raphson method (e.g., Press et al. 1985) as
$$\delta T^j(r,\Theta)=\dfrac{\Delta_T^j}{(\partial\Delta_T/\partial
T)^j}=\dfrac{\Delta_T^j}{(\Delta_T^{j-1}-\Delta_T^j)/\delta T^{j-1}}$$
for $\Delta_T^j(r,\Theta)\geq 0.1$ and $\delta T^j(r,\Theta)=0$ for
$\Delta_T^j(r,\Theta)<0.1$, where
$\Delta_T^j(r,\Theta)\equiv\{T^{*j}(r,\Theta)-T^j(r,\Theta)\}/T^j(r,\Theta)$.
(e) Compute $\delta T^j(^{\forall}r,^{\forall}\Theta)$ for all grid
points in the same manner. (f) Return to step (a) with the new tentative
temperature $T^{j+1}(r,\Theta)=T^j(r,\Theta)+\delta T^j(r,\Theta)$ at
each point, and repeat these operations until 
$|\Delta_T^j(^{\forall}r,^{\forall}\Theta)|<0.1$ and
$|T^j(r,\Theta)-T^{j-1}(r,\Theta)|/T^{j-1}(r,\Theta)<0.1$ for
$(^{\forall}r,^{\forall}\Theta)$.

(III) Making use of $T^i(^{\forall}r,^{\forall}\Theta)$, compute the
density distribution $\rho^i(^{\forall}x,^{\forall}z)$ from equation
(\ref{1}) as we stated in \S 2.1.

(IV) Estimate $T^{i+1}(r,\Theta)$ in the optically
thick region by using $\rho^i(^{\forall}x,^{\forall}z)$ and solving
equation (\ref{12}) as we stated in the process (I). This is because
we can hardly get the adequate temperature from a wrong initial value of
$T^{j=1}(r,\Theta)$ in very optically thick region, owing to the
approximate treatment of integral (\ref{10}) and to the determination of
the temperature correction $\delta T^j(r,\Theta)$ in (II). As we
describe in \S A.2, the Short Characteristics method to integrate
equation (\ref{10}) lets $I_{\nu}(r,\Theta;\mu,\phi)\rightarrow
B_{\nu}[T(r,\Theta)]$, which leads to $\Delta_T^j(r,\Theta)\rightarrow
0$ and then $\delta T^j(r,\Theta)\rightarrow 0$, for very optically thick
limit. (This is also why we choose the initial density and temperature
distributions as [I].)

(V) Return to step (II) until
$|\rho^i(x,z)-\rho^{i-1}(x,z)|/\rho^{i-1}(x,z)<0.1$ and
$|T^i(r,\Theta)-T^{i-1}(r,\Theta)|/T^{i-1}(r,\Theta)<0.1$ for 
$(^{\forall}x,^{\forall}z)$ and $(^{\forall}r,^{\forall}\Theta)$.

\subsection{Short Characteristics Method to Calculate Intensity}

Next we mention the numerical method to integrate equation (\ref{10}) in
order to get the specific intensity $I_{\nu}(r,\Theta;\mu,\phi)$. Since
it costs vast CPU time to compute the integration from $s'=0$ to $s$ in
all directions $(\mu,\phi)$ and all frequencies $\nu$, we here adopt an
efficient and convenient algorithm, the Short Characteristics method in
the spherical coordinates (Dullemond \& Turolla 2000; see also Mihalas,
Auer, \& Mihalas 1978, Olson \& Kunasz 1987, Auer \& Paletou 1994).
Following them, we approximate to calculate the intensity at a
point $P$ under the assumption of neglecting the scattering (that is,
only the black body radiation $B_{\nu}[T(r,\Theta)]$ is considered for
source function at each point) as
\begin{equation}
I_{\nu}(P;\mu,\phi)\approx e^{-\tau_{\nu}}I_{\nu}(U;\mu,\phi)+u_{\nu}B_{\nu}(U)+p_{\nu}B_{\nu}(P)+d_{\nu}B_{\nu}(D), \label{20}
\end{equation}
where $U$ and $D$ represent the upstream and downstream points along a
ray directed to $(\mu,\phi)$, and $\tau_{\nu}$ is the optical depth
between the points $U$ and $P$. The coefficients $u_{\nu}$, $p_{\nu}$,
and $d_{\nu}$ are functions of the optical depths from $U$ to $P$ and
from $P$ to $D$
(see e.g.,
Dullemond \& Turolla 2000 for detailed determination of the coefficients).
The intensity (\ref{20}) approaches $I_{\nu}(P;\mu,\phi)\rightarrow
I_{\nu}(U;\mu,\phi)$ for optically thin limit, $\tau_{\nu}\rightarrow 0$,
and $I_{\nu}(P;\mu,\phi)\rightarrow B_{\nu}(P)$ for optically thick limit,
$\tau_{\nu}\rightarrow \infty$. Making use of this method, we
can obtain $I_{^{\forall}\nu}(^{\forall}
r,^{\forall}\Theta;^{\forall}\mu,^{\forall}\phi)$ systematically from 
the intensity at boundaries (see below) and the temperature at each
point $T(^{\forall}r,^{\forall}\Theta)$.

In this calculation we use the spatial grid of $(r_k,\Theta_l)$,
where $k=1,40$ and $l=1,80$. In order to resolve the inner region of the
disk, the radial grid is put logarithmically as $r_{k+1}=pr_{k}$,
where $p$ is a constant, determined as $r_1=r_{\rm in}$ and
$r_{40}=r_{\rm out}$ are satisfied. We assign the inner radius as $r_{\rm
in}=R_*=2R_{\odot}$ and the outer radius as $r_{\rm out}=100$AU.
The azimuthal grid points are equally spaced in $0<\Theta<\pi/2$. 
For the direction of the ray-path $(\mu,\phi)$, we put 32 points at the
maximum for $\mu$ in $0\leq\mu\leq1$ (see Dullemond \& Turolla 2000 in
detail for choosing the $\mu$-grid), and equally spaced 8 points for
$\phi$ in $0<\phi<\pi/2$. The frequency is divided logarithmically into
32 points in $10^{10}\leq\nu\leq 10^{15}$ Hz. 
The boundary conditions that we adopt here are as follows:
$I_{\nu}(r=r_1,\Theta;\mu>0,\phi)=B_{\nu}(T=T_*=4000{\rm K})$ is taken
for the intensity to the outward direction at the inner radius $r=r_1$,
and $I_{\nu}(r=r_{40},\Theta;\mu<0,\phi)=B_{\nu}(T=10{\rm K})$ 
for the inward intensity at the outer radius $r=r_{40}$ 
(the gaseous matter in molecular clouds is considered to have the
equilibrium temperature of about $T=10$K between heating by cosmic rays
and cooling by molecular line emissions; e.g., Myers 1978).
At $\Theta=\Theta_{1}$ we take the symmetric condition against 
$z$-axis ($\Theta=0$), and at $\Theta=\pi/2$ we put 
$I_{\nu}(r,\Theta=\pi/2;\mu,\phi)=B_{\nu}(T=T_{\rm vis})$.
We note that the boundary conditions at $r=R_*$ and $\Theta=\pi/2$
reproduce the heating sources, the central star and the viscous
dissipation at the midplane, respectively, that we described in \S 2.2
and \S 2.3.


\clearpage
\begin{figure}
\vspace{-1cm}
\plotone{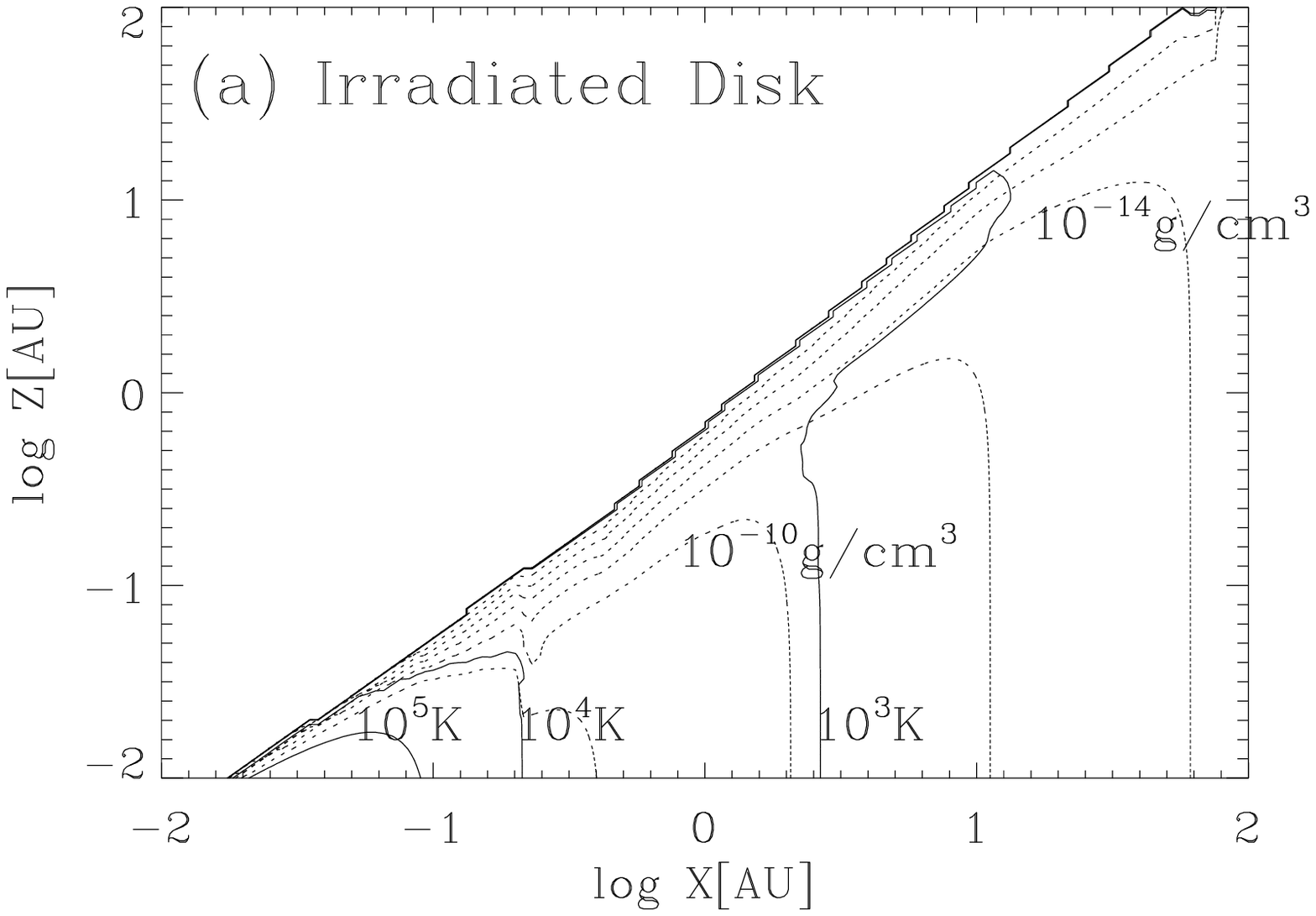}
\vspace{-1cm}
\plotone{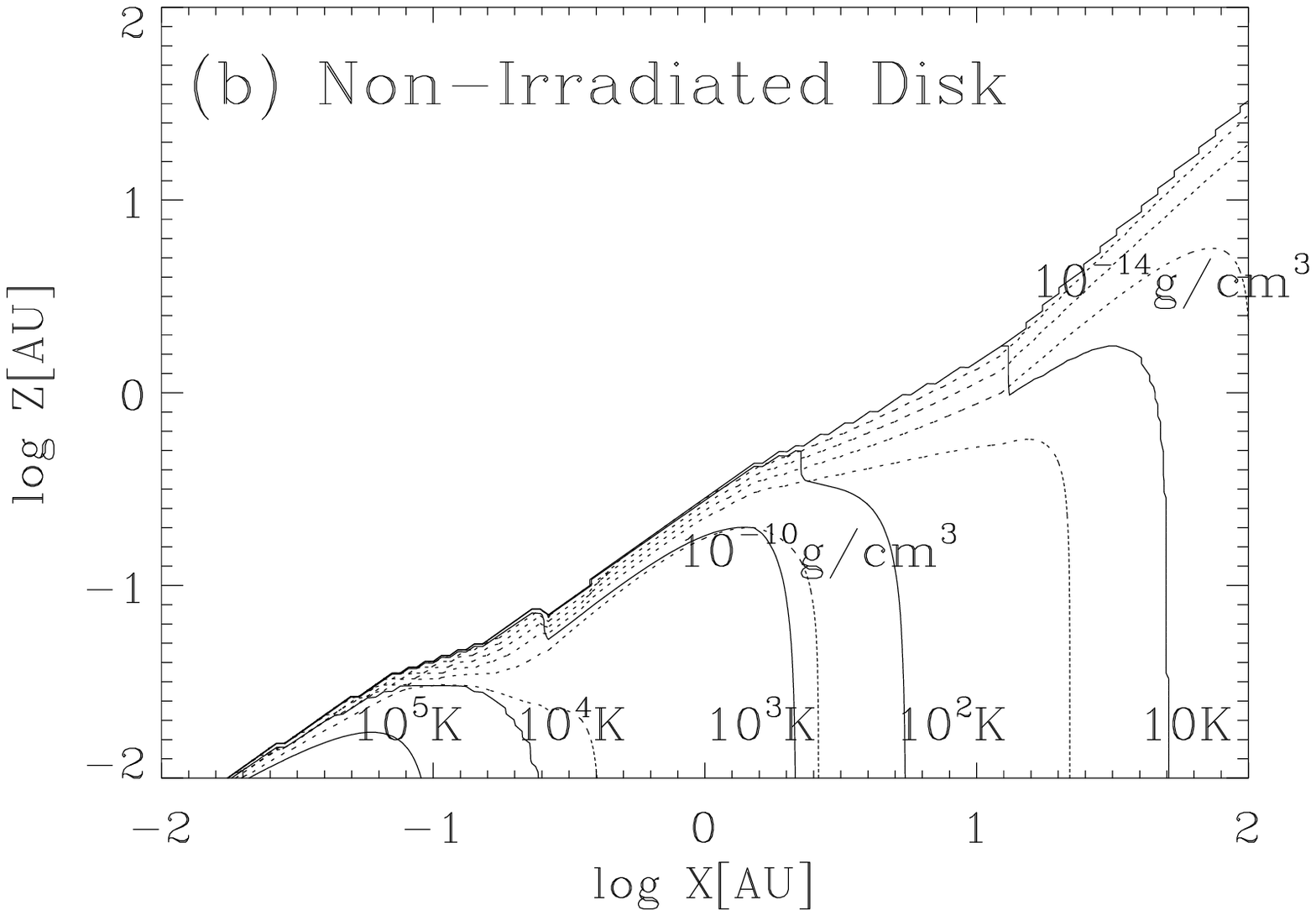}
\caption{The temperature ({\it solid lines}) and density ({\it dotted
 lines}) distributions in the $x$-$z$ plane of (a) `irradiated disk'
 (obtained from 2-D radiative transfer calculation) and (b)
 `non-irradiated disk' (derived from only vertical energy transfer
 equation). The  contour levels are 
 $T=10,10^2,10^3,10^4$, and $10^5$K for the temperature, and 
 $\rho=10^{-18},10^{-16},10^{-14},10^{-12},10^{-10}$, and $10^{-8}$g cm$^{-3}$ 
 for the density. The temperature distributions at the surface region
 and at the midplane outside $x\sim 1$AU and the disk height of the
 `irradiated disk' are higher than those of the `non-irradiated disk'.}
\end{figure}

\clearpage 

\begin{figure}
\vspace{-0.5cm}
\plotone{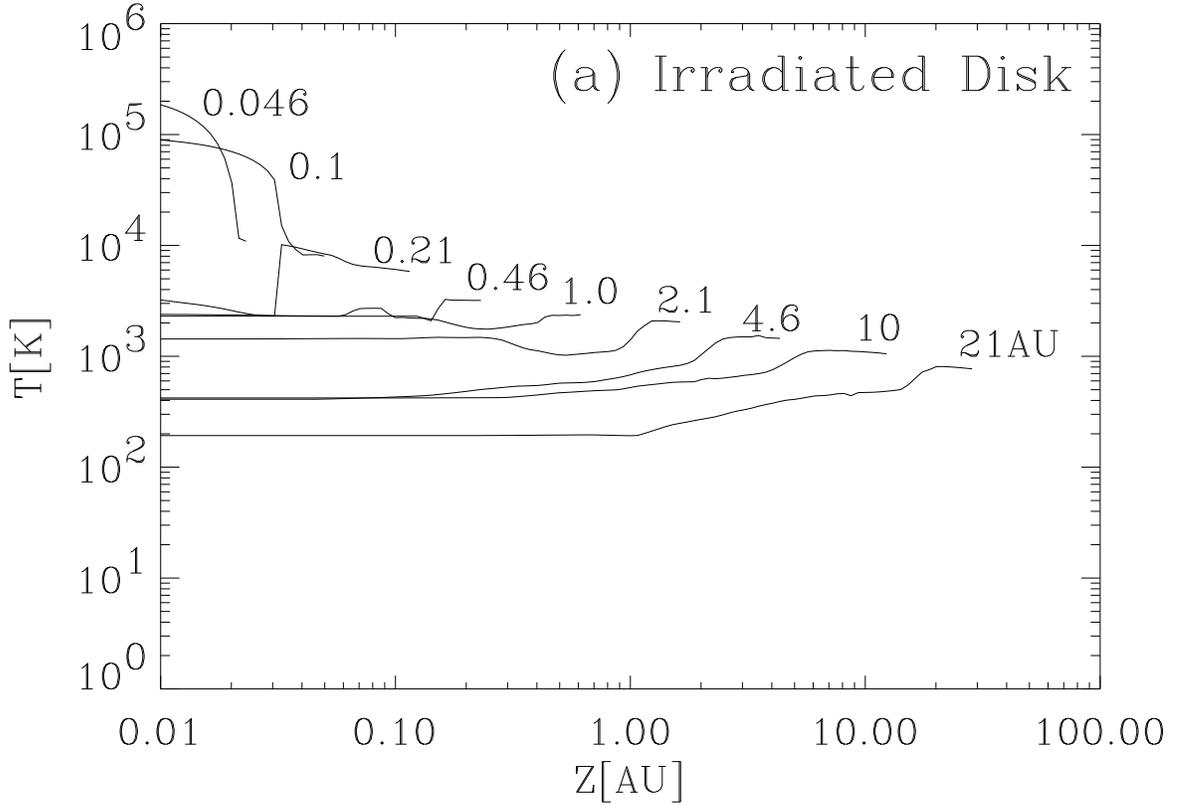}
\vspace{-0.5cm}
\plotone{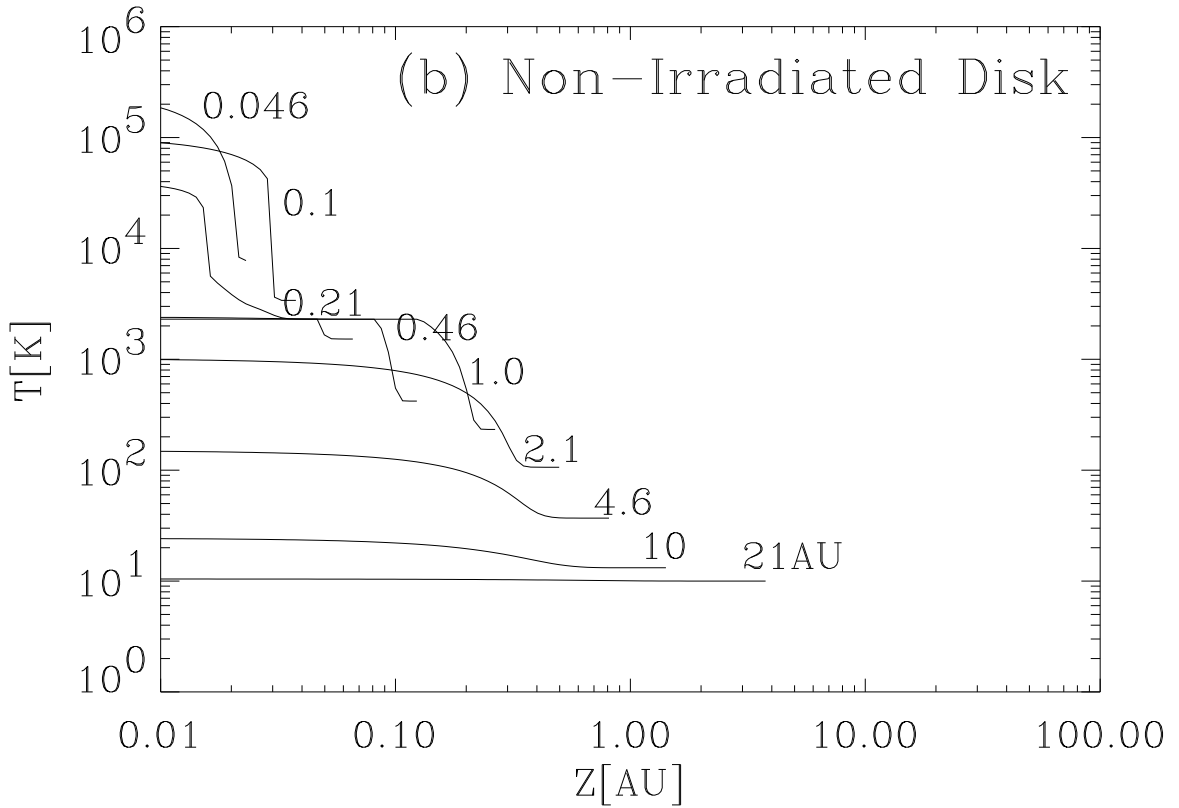}
\caption{The vertical temperature profiles at
 $x=0.046,0.1,0.21,0.46,1.0,2.1,4.6,10$, and $21$AU of (a) `irradiated 
 disk' and (b) `non-irradiated disk'.}
\end{figure}

\clearpage 

\begin{figure}
\vspace{-0.5cm}
\plotone{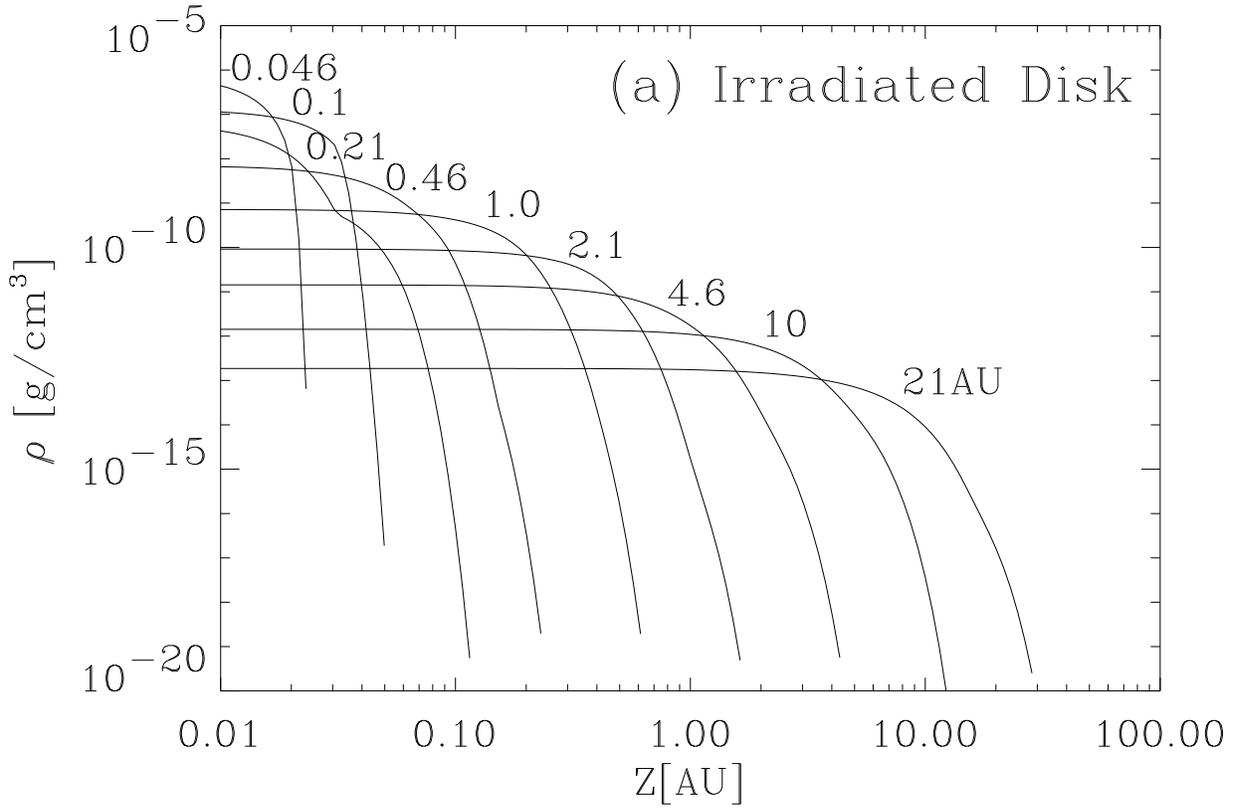}
\vspace{-0.5cm}
\plotone{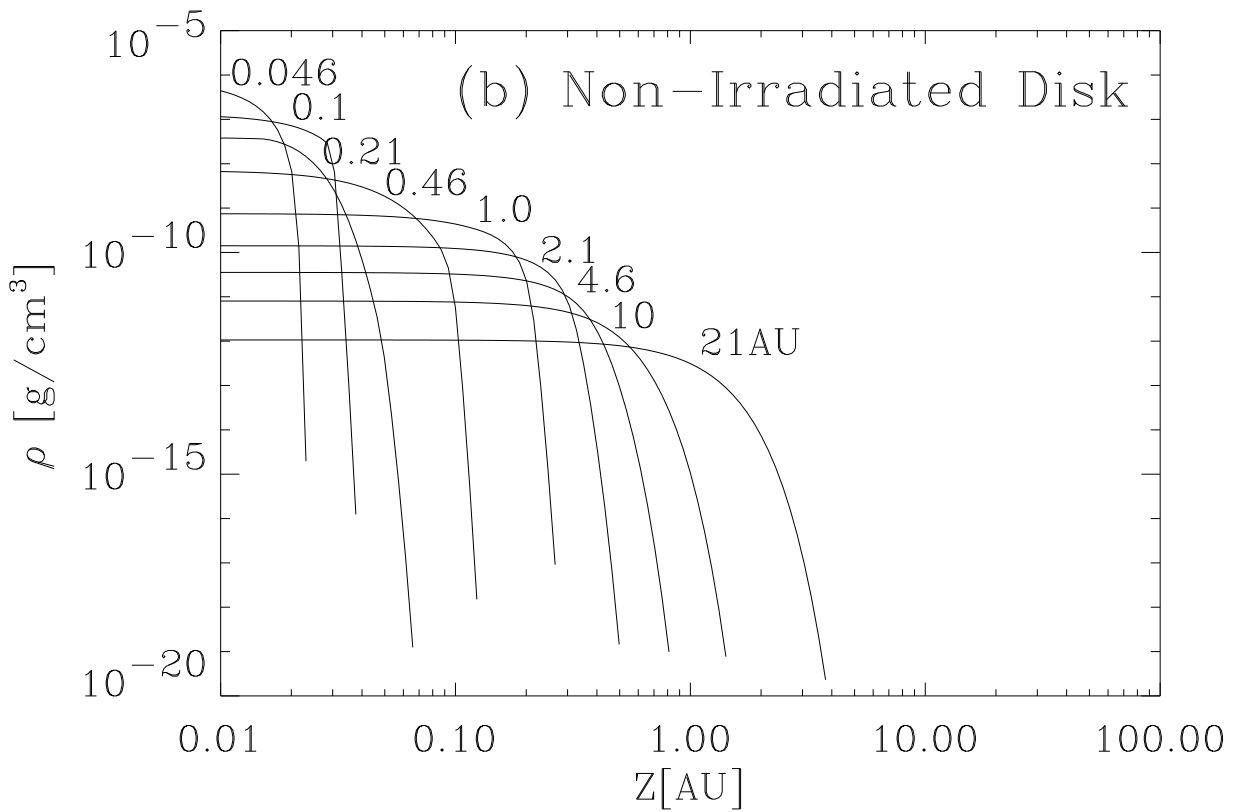}
\caption{The same as Figure 2 but the vertical density profiles.}
\end{figure}

\clearpage 

\begin{figure}
\vspace{-0.5cm}
\plotone{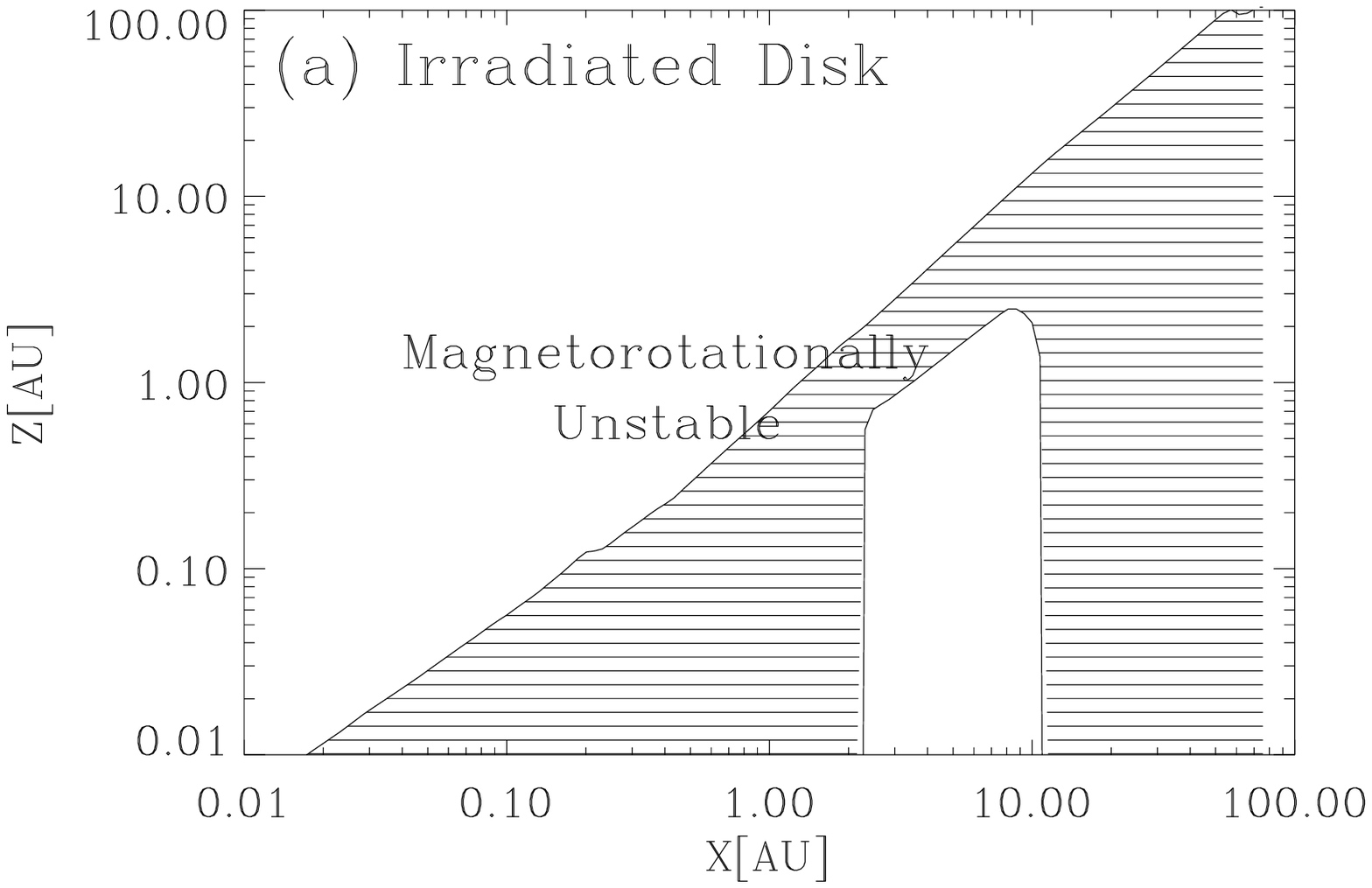}
\vspace{-0.5cm}
\plotone{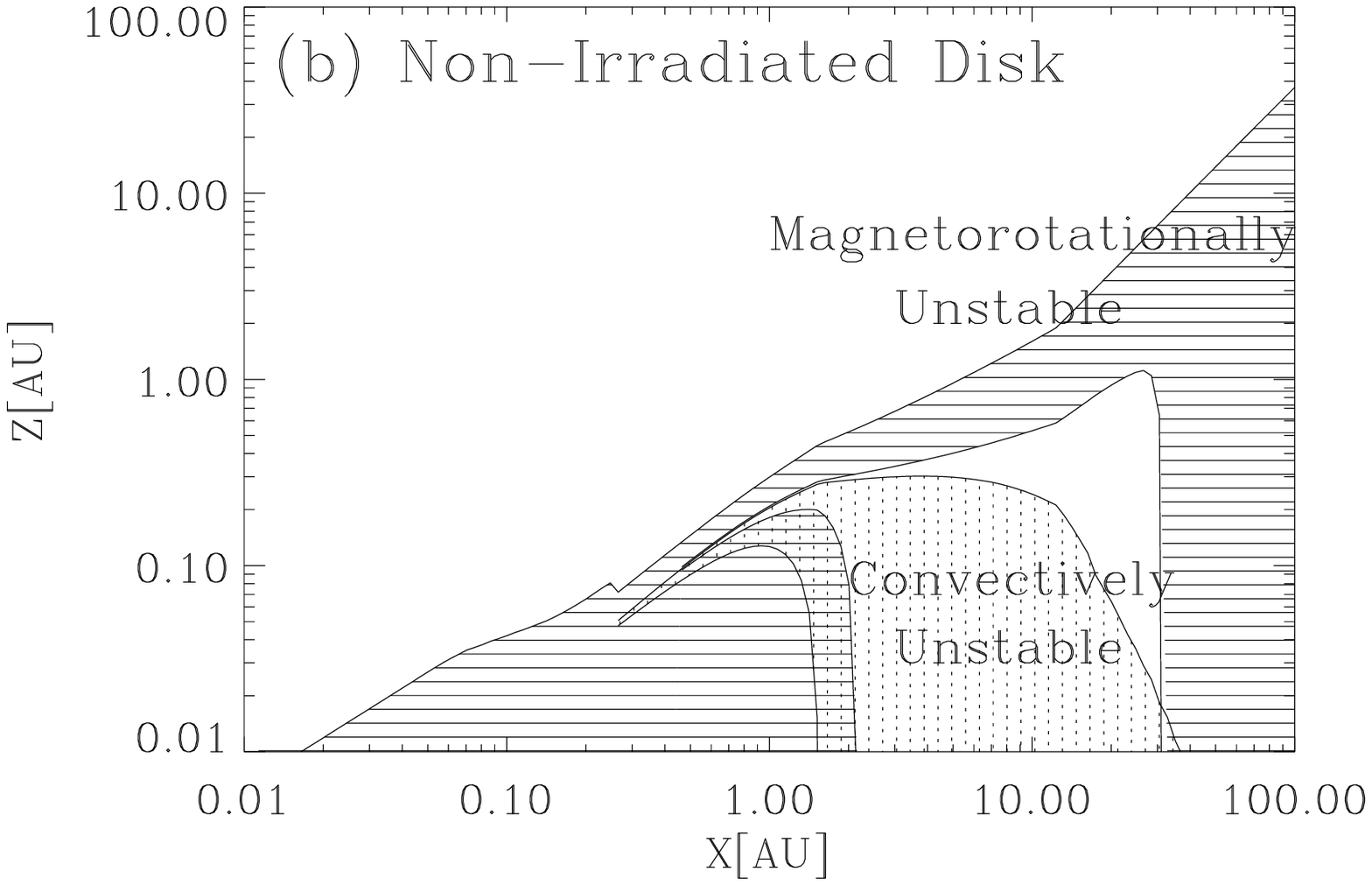}
\caption{The convectively ({\it dotted stripe}) and the 
 magnetorotationally ({\it solid stripe}) unstable regions  in the
 $x$-$z$ plane of (a) `irradiated disk' and (b) `non-irradiated
 disk'. The whole `irradiated disk' is convectively stable, while
 unstable region exists in the `non-irradiated disk'. Almost all the
 `irradiated disk' is magnetorotationally unstable except for the region
 near the equatorial plane of 2AU $<x<$ 10AU.}
\end{figure}

\clearpage 

\begin{figure}
\plotone{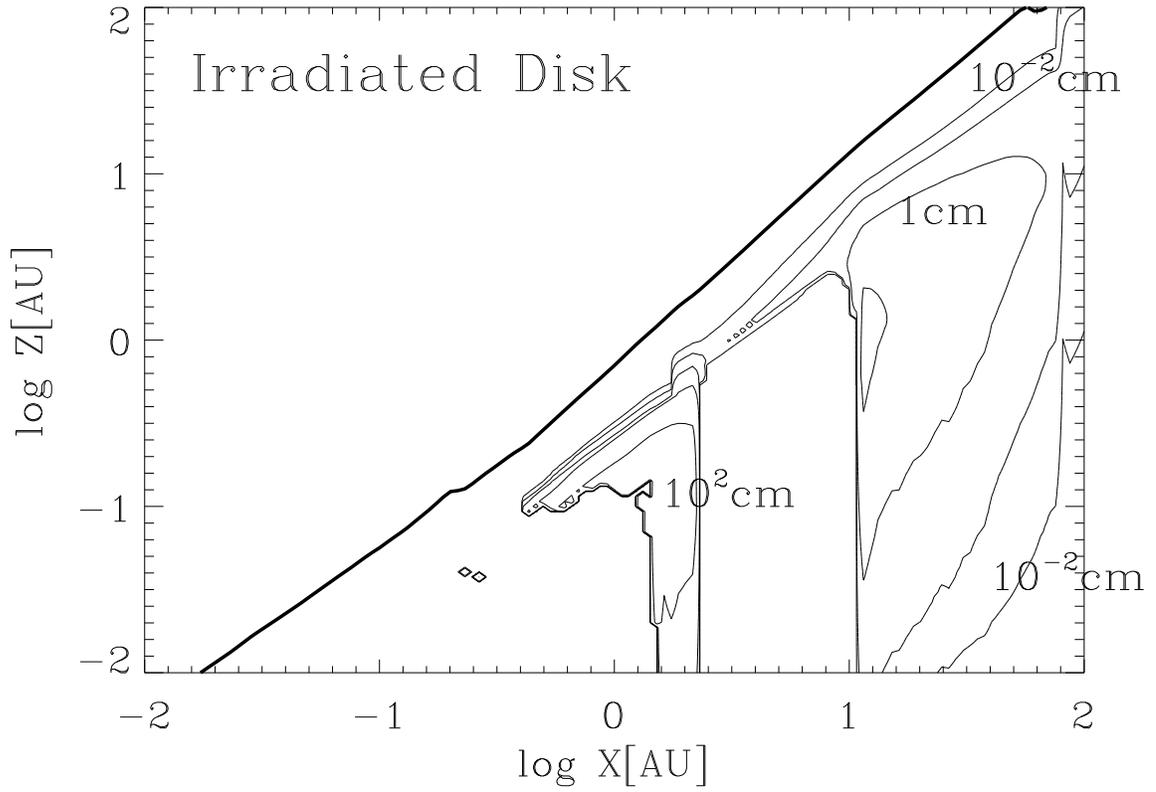}
\caption{The maximum dust size distribution in the $x$-$z$ plane of
 `irradiated disk'. The contour levels are 
 $a_{\rm p,max}=10^{-2},10^{-1},1,10$, and $10^2$cm. The thick line
 displays $z=z_{\infty}$. The dust particles can grow to be cm-size in
 the surface layer, which are expected to be under the influences of the
 outside activities.} 
\end{figure}

\end{document}